# Hidden Antipolar Order Parameter and Entangled Néel-Type Charged Domain Walls in Hybrid Improper Ferroelectrics

M. H. Lee,[1] C.-P. Chang,[1] F.-T. Huang,[2] G. Y. Guo,[3,4] B. Gao,[2] C. H. Chen,[1,2] S.-W. Cheong,[2] and M.-W. Chu[1*]

[1]*Center for Condensed Matter Sciences, National Taiwan University, Taipei 106, Taiwan*

[2]*Rutgers Center for Emergent Materials and Department of Physics and Astronomy, Rutgers University, Piscataway, New Jersey 08854, USA*

[3]*Department of Physics, National Taiwan University, Taipei 106, Taiwan*

[4]*Physics Division, National Center for Theoretical Sciences, Hsinchu 300, Taiwan*

[*]Corresponding author: M.-W. Chu (chumingwen@ntu.edu.tw)

Hybrid improper ferroelectricity (HIF) denotes a new class of polar instability by the mixture of two octahedral-distortion modes and can feature the coexistence of abundant head-to-head and tail-to-tail polar domains, of which the domain walls tend to be charged due to the respective screening charges with an opposite sign. However, no such coexisting carriers are available in the materials. Using group-theoretical, microscopic, and spectroscopic analyses, we established the existence of hidden antipolar order parameter in model HIF $(Ca,Sr)_3Ti_2O_7$ by the condensation of a weak, previously unnoticed antipolar lattice instability, turning the order-parameter spaces to be multicomponent with the distinct polar-antipolar intertwining and accompanied formation of Néel-type twin-like antipolar domain walls (few nm) between the head-to-head and tail-to-tail domains. The finite-width Néel walls and correlated domain topology inherently lift the polar divergences between the domains, casting an emergent exemplification of charged domain-wall screening by an antipolar ingredient. Comparisons to topological defects in improper-ferroelectrics hexagonal manganites were discussed.

PACS indexing codes:  77.80.Dj   61.50.Ks   61.14.Lj   79.20.Uv

The macroscopic Landau theory of phase transitions depicts the grand fundamental of a plethora of phenomena ranging from ferroelectricity [1-3] to density waves [4], with the ferroelectric (FE) transition being the textbook example for general structural phase transitions in solids [5,6]. In the corresponding group-theoretical context, proper FEs refers to a material with the spontaneous polarization as primary order parameter, which transforms like a zone-center polar lattice instability, and improper FEs delineates ferroelectricity induced by order parameter belonging to a zone-boundary non-polar irreducible representation (irrep) with FE polarization being the secondary order parameter upon the transition [1,2,5-7].

In oxides, the zone-boundary instability can be linked to oxygen polyhedral distortions [7-9]. The general antiferrodistortive octahedral buckling in $ABO_3$ perovskites (A and B, respective 12- and 6-fold coordinated cations; O, oxygen) is particularly intriguing considering the two ubiquitous, yet competing, order parameters of zone-center FE and zone-boundary octahedral instabilities in the bulks [6,10-12]. Further upon heterojunction, the translational symmetry generic to the bulks is broken across the interface and the heterostructure can be subject to misfit strain, perturbing the existing order-parameters competition and likely mediating two-dimensional interfacial phenomena [13,14]. The rejuvenated FE instabilities in $LaAlO_3/SrTiO_3$ heterojunctions ($LaAlO_3$ and $SrTiO_3$, nominally free from FE ordering) [13] and the improper ferroelectricity in $PbTiO_3/SrTiO_3$ superlattices ($PbTiO_3$, originally proper FEs) [14] are exemplifications of such two-dimensional engineering.

Indeed, Ruddlesden-Popper oxides, $(AO)$-$(A_nB_nO_{3n})$, naturally crystallize into two-dimensional perovskites (n, perovskite-unit number) [15], with the rock-salt AO layer sectioning the three-dimensional corner-shared octahedra into two-dimensional perovskite slabs that nurture more octahedral degrees of freedom [15,16]. The Ruddlesden-Popper phases hence display rich octahedral distortions [15,16] and the recently coined hybrid improper ferroelectricity (HIF) in $n = 2$ $Ca_3Ti_2O_7$ (CTO) and $(Ca,Sr)_3Ti_2O_7$ is particularly enlightening with the zone-center $\Gamma$-point ferroelectricity being driven by the hybrid condensation of two zone-boundary octahedral instabilities that transform like two-dimensional *X*-point irreps [17-22].

Macroscopically, the CTO and (Ca,Sr)$_3$Ti$_2$O$_7$ are distinguished from prototypical proper FE BaTiO$_3$ [23-25] and improper FE rare-earth molybdates [1-3,7-9] by the abundant head-to-head (HH) and tail-to-tail (TT) domains, where the FE dipoles point toward and away from each other across the respective domain walls (DWs) [18-20]. Accordingly, notable electrostatic divergences arise therein and the HH and TT domains cannot be stable without the complementary screening charges at the DWs, while only electrons are available in n-type titanates [13,18,23-25]. The microscopic screening at the HH and TT DWs in CTO and (Ca,Sr)$_3$Ti$_2$O$_7$ is thus highly interesting and remains unsettled despite the proposal of topological protection by antiphase boundaries (APBs) [19,20]. Here, we report the atomic-scale observation of finite HH and TT DWs (few nm, width) with an unanticipated antipolar structure in model HIF Ca$_{2.5}$Sr$_{0.5}$Ti$_2$O$_7$ (CSTO) and the correlated screening of the polar divergences using macroscopic group-theoretical analysis and microscopic structural and electronic investigations by (scanning) transmission electron microscopy, (S)TEM, and electron energy-loss spectroscopy (EELS). This domain topology and underlying group-theoretical principles refine the understanding in physics of complex FE domains [19,20,26,27].

Figs. 1a and 1b exhibit the FE ground-state structure (space-group *A2$_1$am*) along respective *b′* and *a′* projections (*a′* ~ *b′* ~ √2*a*, *c′* ~ *c* thus *c* for simplicity; *a* and *c*, parent tetragonal lattice parameters) and the FE polarization (order parameter, *P*) along *a′* axis [20]. Using the point-charge approximation for FE-dipole estimations [13], we derived each atomistic contribution to *P* in an individual perovskite slab (gray region, Fig. 1a) in Fig. 1c. Fig. 1d represents the group-theoretical analysis of the symmetry tree [28] for transition pathways.

Compared to the paraelectric parent phase (*I4/mmm*), the antiferrodistortive octahedral tilting (order parameter, *T*) in *a′b′*-plane in Figs. 1a and 1b originates from the lattice instability against $X_3^-$ irrep and the order parameter of octahedral rotation (*R*) along *c* axis represents the $X_2^+$-irrep distortion mode, altogether known as the hybrid condensation of the two *X*-point irreps [17-22]. This cooperative transition lifts any single direct route to *A2₁am*, with the polar *P* addressed by the zone-center $\Gamma_5^-$ irrep (Fig. 1d) [17,21,29]. It is noted that the zone-boundary $Z_4$ link for the pathway from $\Gamma_5^-$-induced *F2mm* to *A2₁am* (Fig. 1d) renders the resultant *A2₁am* nonferroic [5], ruling out the FE ground state as a child group of $\Gamma_5^-$ irrep. This latter feature confirms the HIF notion of *P* as the product of *R* and *T* [17,21] and the ferroelectricity in CSTO was readily ascribed to the $\Gamma_5^-$-induced antiparallel Ca1/Sr1 and Ca2/Sr2 displacements along *a′* axis (white arrows, Fig. 1a) [18-20,30]. The atomistic decomposition of *P* (Fig. 1c), however, unveils that all $\Gamma_5^-$-related Ca/Sr, Ti, and O displacements along *a′* are involved [21,22]. The thus-derived *P* of ~2.14 μC/cm$^2$ for a perovskite slab, i.e., ~4.28 for an unit cell (uc), is compatible with the measured ~2.97 μC/cm$^2$ in single crystals [18]. Fig. 1c hence suggests that the formed crystallographic opinions on the HIF [17,18,29,30] deserve further elaborations. The nominal antipolar Ca2/Sr2 displacements along *b′* axis (gray arrows, Fig. 1b), neglected before [20], provide a useful hint.

Figure 2a shows the *b′*-projected STEM high-angle annular dark-field (HAADF) image of HH domains. Fig. 2b exhibits one set of STEM-EELS chemical maps, exploiting Ca-$L_2$, Sr-$M_3$, and Ti-$L_2$ edges [13,31]. Fig. 2c depicts the characteristic HAADF imaging along *a′* projection. The *b′*-projected DF TEM imaging of different specimen regions are shown in Fig. 2d. Fig 2e represents the TT counterpart to Fig. 2a. Each panel in Fig. 2 was acquired in crystalline areas well away from any twin boundaries and, therefore, denotes the inherent structural characteristics.

Compared to pristine CTO, the Sr substitution increases the domain density by accompanied reduction in the *a′b′*-orthogonality and related ferroelastic-strain cost, rendering DW investigations convenient with various DW angles [18-20,31]. The larger, heavier Sr preferentially occupies the spacious perovskite A site (Fig. 2b; Sr map, blue) compared to the 9-fold rock-salt A site (Ca map, red), without introducing additional distortion to the CTO [15,18] and accounting for the enhanced Ca1/Sr1 contrasts in the atomic-number sensitive HAADF imaging (red rectangles, Fig. 2a).

A careful examination of Fig. 2a unveils that the HH DW (yellow) shows a different structure from that in the neighboring domains. Surprisingly, the DW structure mimics the $a'$-projected CSTO (Fig. 2c), with the $b'$-oriented antipolar Ca2/Sr2 displacements in Fig. 1b being attenuated (red-margined arrows, bottom-left inset) and the nominally quenched Ca1/Sr1 rejuvenated and exhibiting antipolar distortions (white-margined arrows). These accentuated $b'$-oriented antipolar Ca/Sr displacements lead to the previously unnoticed distortion of hourglass- and barrel-like perovskite units along $c$-stacking (Fig. 2c). The TT DW (yellow, Fig. 2e) shows the same distortion pattern of hourglass-barrel stacking as the HH counterpart (yellow, Fig 2a) and $a'$-projected CSTO (Fig. 2c). Fig. 2d reveals that the HH and TT domains are equally populated. In Figs. 3 and 4, we scrutinize the antipolar displacements as hidden order parameter and its role in screening the HH and TT domains.

Figure 3a shows the calculated phonon dispersion of CTO that exhibits identical antipolar distortion pattern (inset) to CSTO [31]. The negative frequency and local minimum at a given reciprocal lattice point indicate the uc instability against the irrep [13,38]. Fig. 3a thus reveals the existence of $\Gamma$-, $X$-, and $P$-point soft phonons [1-3,6-8,16,38], with the $N$- and $Z$-point saddles arising from the proximity to $\Gamma$- and $P$-point instabilities considering their incompatibility with the symmetry tree (Fig. 1d) [5,32,39]. The pronounced $X$- and $P$-point dips in Fig. 3a signify their important roles in the ground-state structure [38] and the shallower $\Gamma$-point phonon is consistent with the HIF by $X$-point instabilities [17,22]. Notably, $P$-point instability is undocumented in the earlier theoretical [17,22,29,30] and X-ray and neutron powder studies of the HIF [15,20,21], while admissible for Ruddlesden-Popper phases [16].

The thermal diffused scattering in convergent-beam electron diffraction (CBED), owing to electron-phonon interaction, is a fundamental map of symmetry elements within the phonon spectrum [40]. CBED is then complementary to X-ray and neutron powder diffractions when probing intricate structural distortions is limited by the diffraction peak-intensity and -overlap subtleties [15,20,21,46]. Figs. 3b-3d show the $b'$-, $a'$-, and $c$-projected CBED patterns of CSTO, showing Bragg-scattered discs with dynamical-interference fringes in the bright field (BF; center, transmitted disc) and striped thermal-diffused Kikuchi bands in the whole pattern (WP). A careful examination of the BF in Fig. 3b reveals the characteristic absence of a mirror perpendicular to $a'^*$ and the Kikuchi bands in the WP (green stripes) map the *2mm* point-group symmetry of $A2_1am$ by the two perpendicular mirrors (white) [15]. Likewise, the BF-*2mm* and WP-*2mm* symmetries in Fig. 3c agree with the $a'$-projected *2mm* point group of $A2_1am$ [15]. Surprisingly, the $c$-projected Fig. 3d shows *2mm* BF and *2*-fold WP considering the absent mirror operation between group-1 Kikuchi bands and group-2 and -4 counterparts (blue stripes; guiding white, red arrows) and the *2*-fold operation for groups 1 and 3. This BF-WP symmetry combination leads to $2m_Rm_R$ diffraction group that corresponds to *222* point group [47].

Figs. 3b-d thus suggest that there exists a weak *222*-type distortion. An investigation of the isotropy subgroups of $I4/mmm$ reveals that point-group *222* is bound to the *P*-point irrep of $P_5$, with the $P_5$-irrep $F222$ (Fig. 1d) allowing the $b'$-oriented antipolar Ca1/Sr1 and Ca2/Sr2 displacements in Figs. 2 and 3a (inset) [5,39]. The condensation of *P*-point instability (Fig. 3a) is unambiguously correlated with the $P_5$-irrep antipolar distortion, establishing the group-theoretical footing of the antipolar displacements as hidden order parameter. Along $a'$ and $b'$ projections, the weak $P_5$-irrep (1/2, 1/2, 1/2) modulation is likely masked by the (1/2, 1/2, 0) of prominent $X_2^+$ and $X_3^-$, since the long $c$-axis (19.6215 Å; respective $a'$ and $b'$, 5.4362 and 5.4487 Å [20]) brings the $P_5$ modulation closely to the reciprocal $a'^*b'^*$-plane, plausibly explaining why the $P_5$ irrep is observed in Fig. 3d only.

In effect, $P_5$ irrep (Fig. 1d) is composed by two primary order parameters in the basal plane along respective $(a,a)$ and $(b,b)$ directions and two secondary order parameters to be addressed in Fig. 4 [5,16,28,39]. The order-parameter direction of $P_5$ is thus denoted as $(a,a,b,b)$ [5,39], suggesting that the *b'*-oriented antipolar order parameter, i.e., (*b,b*), shall have an *a'*-degenerate counterpart and the physics of CSTO is composed by multicomponent order-parameter spaces of $P_5$, $\Gamma_5^-$, $X_2^+$, and $X_3^-$ (one primary and one secondary order parameters for the latter three two-dimensional irreps), undocumented in the symmetry analyses before [17,22,29,30].

In Figs. 4a-4c, we illustrate the primary *P*, *R*, and *T* order-parameter directions in respective $\Gamma_5^-$, $X_2^+$, and $X_3^-$, and the corresponding four-domain topology [5,39]. Taking Fig. 4a for instance, *P* points along (*a,a*), i.e., *a'* in $A2_1am$ (black square), and is four-fold degenerate with (*a,a*), (-*a,a*), (-*a,-a*), and (*a,-a*) due to the *ab*-degeneracy in *I4/mmm*, casting four domains with the ferroelastic strain (*u*) at DWs (dashed lines) as secondary order parameter [5,28,39]. The in-plane octahedral rotation *R* in Fig. 4b (red arrows) and out-of-plane tilting *T* in Fig. 4c (blue) can be understood likewise [5,30,39]. In $P_5$ (Fig. 4d), an (*a,a*)-oriented antipolar order parameter *A* would nonetheless coincide with *P* (Fig. 4a) and is readily suppressed due to the absent antipolar distortion along *a'* axis (Figs. 1 and 2). The $P_5$ irrep effectively becomes (0,0,*b,b*), with one survived primary order parameter *A* along *b'*, two secondary order parameters of *u* and *P* (Fig. 4e), and four-domain topology considering the reduction from eight by the *ab*-degeneracy [5,39].

Upon the hybrid condensation of *R* and *T*, *P* turns out to be the macroscopic order parameter in the phenomenological domain topology [35,36,48] as well as *A* considering its accompanied observations in Figs. 2a and 2e. The symmetry essences in Figs. 4a-4d are then summarized into Fig. 4e, with the *P* and *A* forming the macroscopic order parameters and being generically intertwined in the four-domain topology (otherwise eight domains upon *P*-direction reversals [18-20,30]). Through this *P-A* pairing, antipolar-*A* twins spontaneously appear between the HH and TT domains (Fig. 4f, plane-view; sandwiched *P*, double headed for arbitrary reversals) by the topology of 1-2-3, 1-4-3, or inherent combination of any three domains in Fig. 4e, with the HH and TT sharing equal probability (indeed observed in Fig. 2d) and the twin-like DWs mimicking the FE Néel walls characterized by an in-plane 90° rotation of the dipole and a finite width (Fig. 4f) [35,36,48,49]. Fig. 4g represents a schematic cross-sectional view of the Néel walls (Fig. 4f) and is experimentally affirmed by Fig. 4h, with the n = 3 intergrowth defect also showing an hourglass-barrel-like distortion within the few-nm DW. This latter feature is in agreement with the generally admissible $P_5$-mode distortion for Ruddlesden-Popper phases [16]. Moreover, it has been theoretically suggested that the emergence of FE Néel walls with finite widths refers to the existence of an additional order parameter within the walls, which can only be allowed in FEs featuring multicomponent order-parameter spaces and is rare in matters [48,49]. This surprising exemplification in CSTO (Figs. 2a, 2e, and 4h) nicely corresponds to this notion by the distinct *P-A* intertwining and accompanied order-parameter spaces (Fig. 4). On either side of the Néel walls (Figs. 4g and 4h), the readily-formed 180°-domain configuration along *c*-stacking leads to coexisting depolarization fields with opposite signs and naturally mitigates the electrostatic divergence thereby, similar to the function of 180° domains in proper FEs [23,35,36,48]. Across the walls (Figs. 4g and 4h), the finite wall width is also helpful in smearing out any residual electrostatic divergence in *a′b′*-plane. Accordingly, the HH and TT DWs are not electrostatically divergent and do not require screening charges as resolved in the STEM-EELS studies in Fig. S1 [31], where an electrostatic-screening essence is excluded. Notably, the FE Néel walls impose a structural screening on the primitively charged DWs and discount the proposed APB-DW characteristics [19,20] considering an APB-based *a′*/2 or *c*/2 translation for the polar-domain unit cells unable to result in the observed hourglass-barrel-like antipolar DW structure.

Indeed, the HH and TT DWs adherent to the six-fold FE vortices in improper-FEs hexagonal manganites represent the first systematically studied domain topology [26,27,34,50] and arise from $Z_6$ topological defects by the trimerized polyhedral tilting characteristic of $K_3$ instability (locked into three phase angles of 0, 2π/3, and 4π/3, i.e., topological $Z_3$ symmetry; accompanied $\varGamma_2^-$ FE degeneracy, $Z_2$; $Z_3 \times Z_2 = Z_6$) [51,52], with the HH and TT DWs being atomically sharp without otherwise structural essence [52-56] and to be electrostatically screened [27,34]. By analogy, the HH and TT DWs in CSTO have been ascribed to $Z_4 \times Z_2$ topological defects ($Z_4$, four-fold degenerate $R$ and $T$; $Z_2$, FE degeneracy; Figs. 4a-d) [19,20], of which the entangled electrostatic screening [27,34] is, however, discarded (Fig. S1) and the characteristic DWs are rather few-nm wide and Néel-type (Figs. 2a, 2e, and 4h). To tackle this subtlety, we performed group-theoretical analysis on the $K_3$ instability in hexagonal manganites and obtained the order-parameter directions of ($a$, 0), (-$a$/2, √3$a$/2), and (-$a$/2, -√3$a$/2), equivalent to the respective $Z_3$ angles of 0, 2π/3, and 4π/3 and being two-fold degenerate with (-$a$, 0), ($a$/2, -√3$a$/2), and ($a$/2, √3$a$/2) like $Z_2$ [39,57]. The analysis also allows domain permutations along the out-of-plane directions as the $Z_6$ topological defects [39,57], altogether suggesting that our phenomenological methodology can be an explicit simple solution to complex domain topologies [48,57], though largely unnoticed before. Future topological-defect elaborations [51,52] by incorporating the *P-A* intertwining shall lead to the same Néel-DW topology as ours, while a dedicated issue on its own.

In summary, the *b′*-oriented antipolar Ca/Sr displacements arise from hidden antipolar order parameter by the condensation of $P_5$ instability. The accepted notion on the ferroelectricity and domains in the HIF has been argued over $\Gamma_5^-$, $X_2^+$, and $X_3^-$ irreps, while insufficient for addressing the antipolar distortion and coexisting HH and TT domains. With the $P_5$ irrep, the order-parameter spaces become multicomponent and the domain topology constitutes intertwined polar and antipolar characteristics, with the sandwiched antipolar Néel-type DWs screening the HH and TT dipoles. The HIF represents a vivid example that structural screening can be an alternative to the conventional electrostatic screening of HH and TT domains. This work unravels the complexity and also flexibility of Ruddlesden-Popper HIF in harboring fascinating physics and would stimulate further studies of structurally-mediated screening in pursuit of new discoveries by thorough group-theoretical explorations in all plausible order-parameter spaces.

**ACKNOWLEDGMENTS**

The authors thank Prof. M. S. Senn (University of Warwick, UK) for insightful discussions. This work was supported by Ministry of Science and Technology, National Taiwan University, and Academia Sinica. The work at Rutgers was funded by the Gordon and Betty Moore Foundation's EPiQS Initiative through Grant GBMF4413 to the Rutgers Center for Emergent Materials. The crystal growth at Rutgers was supported by the NSF MRI grant, MRI-1532006.

**FIGURE CAPTIONS**

FIG. 1. (color online) (a) and (b) The FE-$A2_1am$ crystal structures along $b'$ and $a'$ projections, respectively. The crystallographic sites are indicated (gray, Ca/Sr; cyan, Ti; red, oxygen). White arrows in (a), the antiparallel Ca1/Sr1 and Ca2/Sr2 displacements (black arrow, $P$). Gray arrows in (b), the $b'$-oriented antipolar Ca2/Sr2 displacements. Dashed gray lines in (a) and (b), centered lines for revealing the off-center Ca/Sr displacements. (c) The $a'$-oriented polarization (black) of an individual perovskite slab in (a) and the atomistic contribution of each crystallographic site. (d) Group-theoretical analysis of the symmetry tree, with the black label indicating the space group at each gray-labeled irrep. The symbols in parentheses depict the primary order-parameter directions such as ($a,a$), ($a,a,b,b$), and ($a'$) in corresponding irreps. Solid (dashed) lines, reported (otherwise) transition pathways.

FIG. 2. (color online) (a) The HAADF imaging of HH domains revealing a different feature in the DW (yellow). The Ca1/Sr1 (red rectangle) was used for determining the $P$ direction considering Fig. 1a. (b) The STEM-EELS chemical mapping. Gray (white) circles, Ca/Sr (Ti) omitting the off-center distortions for simplicity. (c) The $a'$-projected HAADF image. Lower-bottom inset, an uc blow-up showing the accentuated antipolar Ca1/Sr1 (white-margined) and Ca2/Sr2 (red-margined arrows) displacements. Dashed white lines, the centered anchors for guiding the eyes. (d) The various DF images (red, blue, and green) exploiting inversed reciprocal-lattice vectors, with the contrast reversal in each set unveiling the domain polarity. (e) The HAADF imaging of TT domains, with the DW structure (yellow) mimicking the HH counterpart in (a) and $a'$-projected (c). $P$ in the DWs and (c), pointing in or out. White rectangles, projected uc.

FIG. 3. (color online) (a) The calculated phonon dispersion of CTO with identical antipolar distortion to the CSTO (inset, $a'$-projected HAADF of CTO). (b), (c), and (d) The CBED patterns along respective $b'$-, $a'$-, and $c$-projections with the BF (gray margined) embedded in the center of the WP. m, mirror. The symmetry characteristics in (b) and (c) refer to the point-group *2mm* of *A2₁am*. In (d), the *2mm* BF and *2*-fold WP (guiding white, red arrows) symmetries suggest point-group *222*-type distortion at $P_5$ irrep. Details, see text.

FIG. 4. (color online) (a), (b), and (c) $\Gamma_5^-$, $X_2^+$, and $X_3^-$ irreps with respective primary order parameters of *P* (black arrows), *R* (red), and *T* (blue) in the basal-plane vector spaces of (*a*,*a*), (0,*a*), and (0,*a*), forming four-domain topologies by the degeneracy labeled on edges. *u* (dashed lines), secondary order parameter of ferroelastic strain at the DWs. Gray (black) uc, *c*-projected parent-tetragonal (FE-orthorhombic) lattice. (d) $P_5$ with effective (0,0,*b*,*b*). Green arrow, primary order parameter of *b′*-oriented antipolar Ca/Sr displacements (*A*, double-headed for the antipolar nature). The eight domains (labels on edges) form a four-domain topology considering the *ab*-degeneracy. (e) Domain topology (upper panel) upon $\Gamma_5^-$-, $X_2^+$-, $X_3^-$-, and $P_5$-irrep condensations. *P* and *A*, intertwined macroscopic order parameters in the four domains (1-4; otherwise eight domains upon *P* reversals). (f) The HH and TT domains with a generically sandwiched antipolar-*A*, Néel-type DW by the inherent domain topology such as 1-2-3 or 1-4-3 in (e). (g) A cross-sectional view of the coexisting HH and TT domains in (f). (h) The experimental HAADF evidence for (g), with the n = 3 defect also showing $P_5$- type antipolar distortion within the DW (yellow). Details, see text.

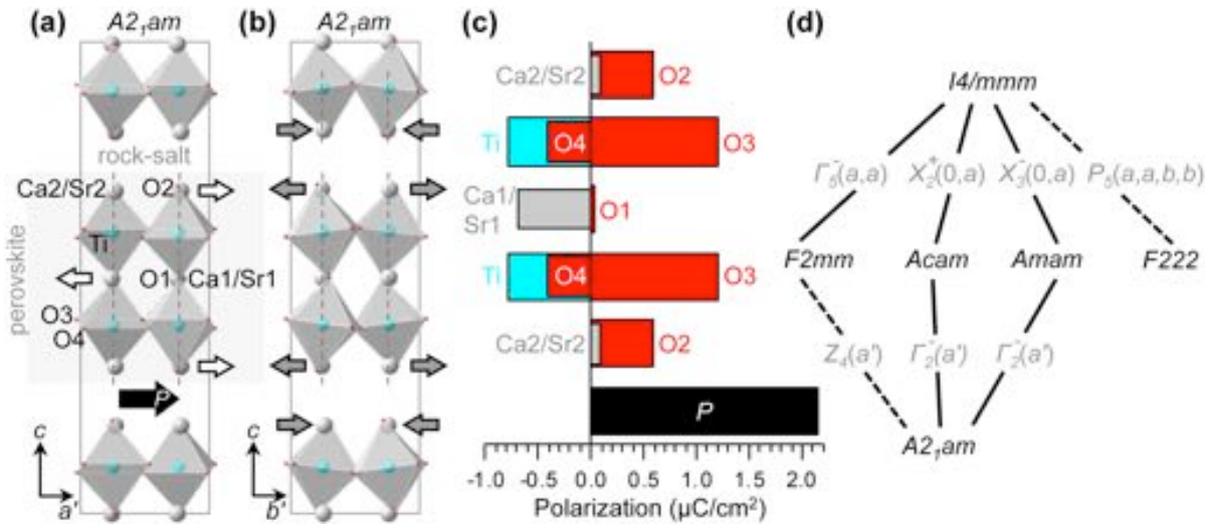

FIG. 1 (M. H. Lee *et al.*)

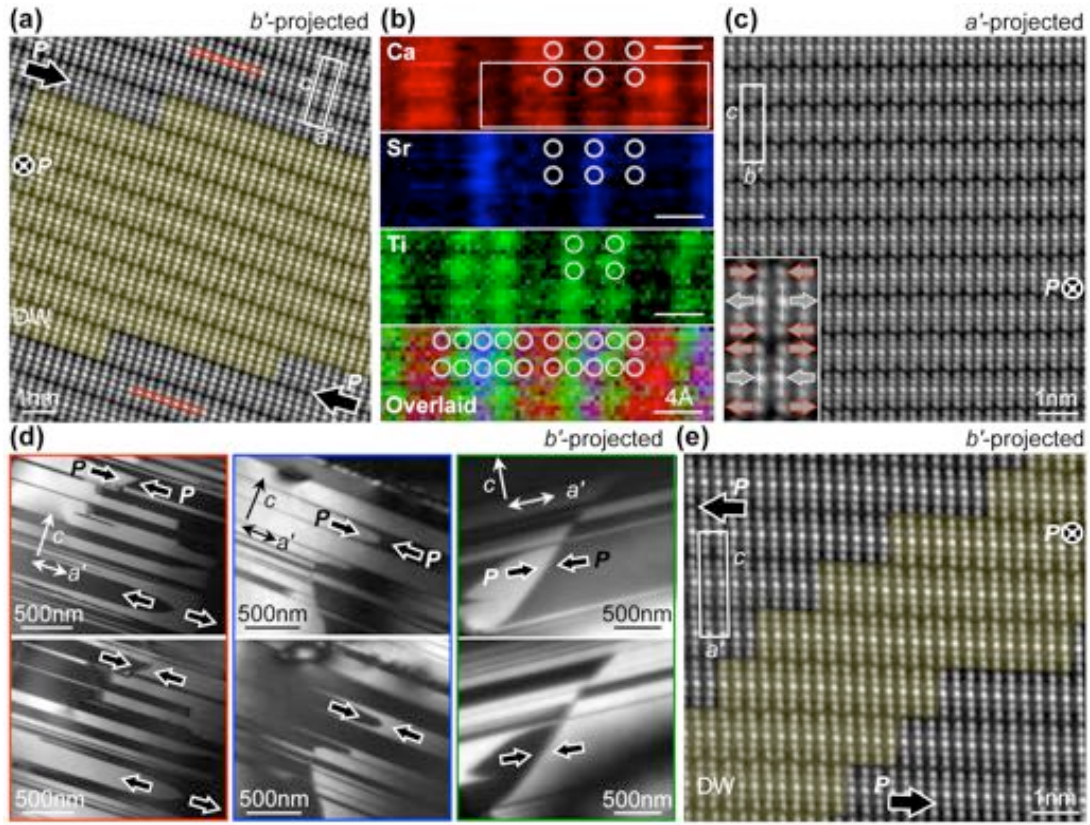

FIG. 2 (M. H. Lee *et al.*)

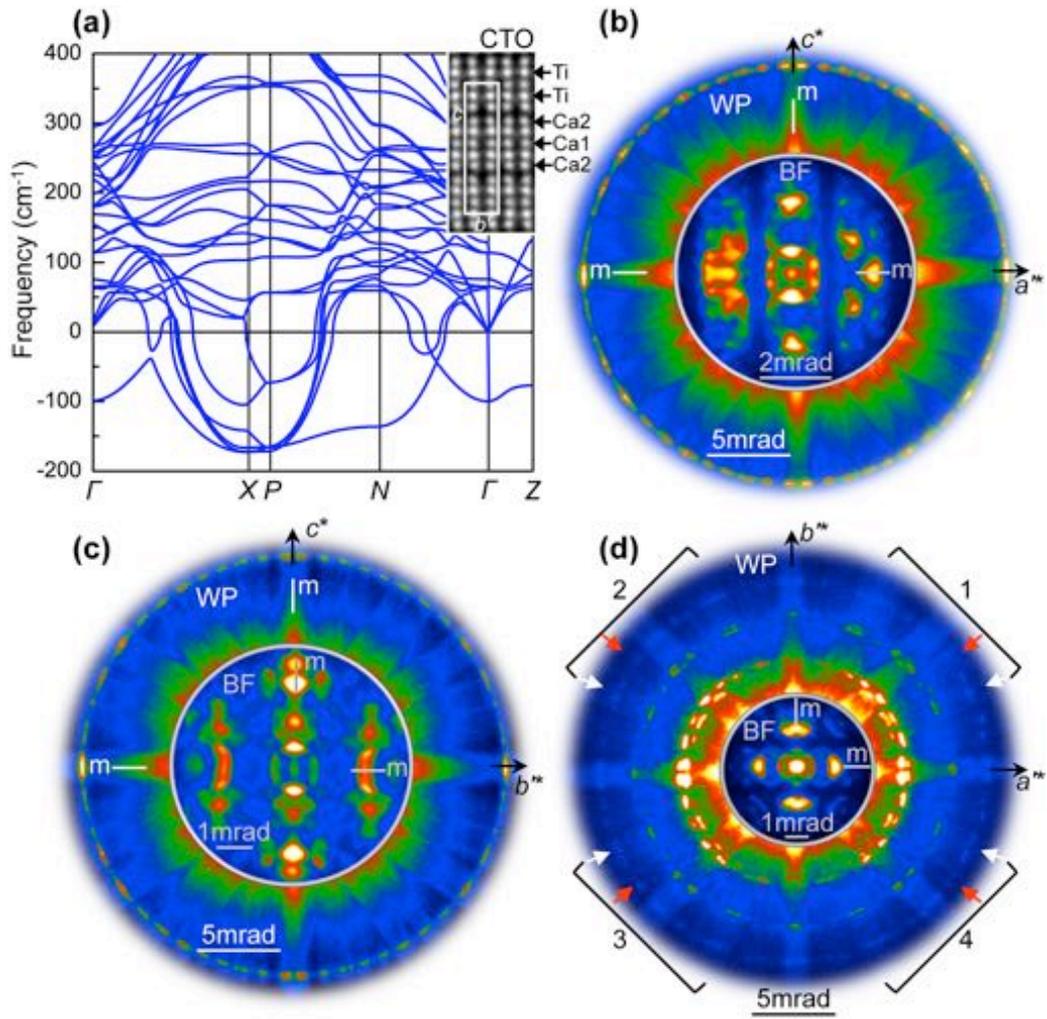

FIG. 3 (M. H. Lee *et al.*)

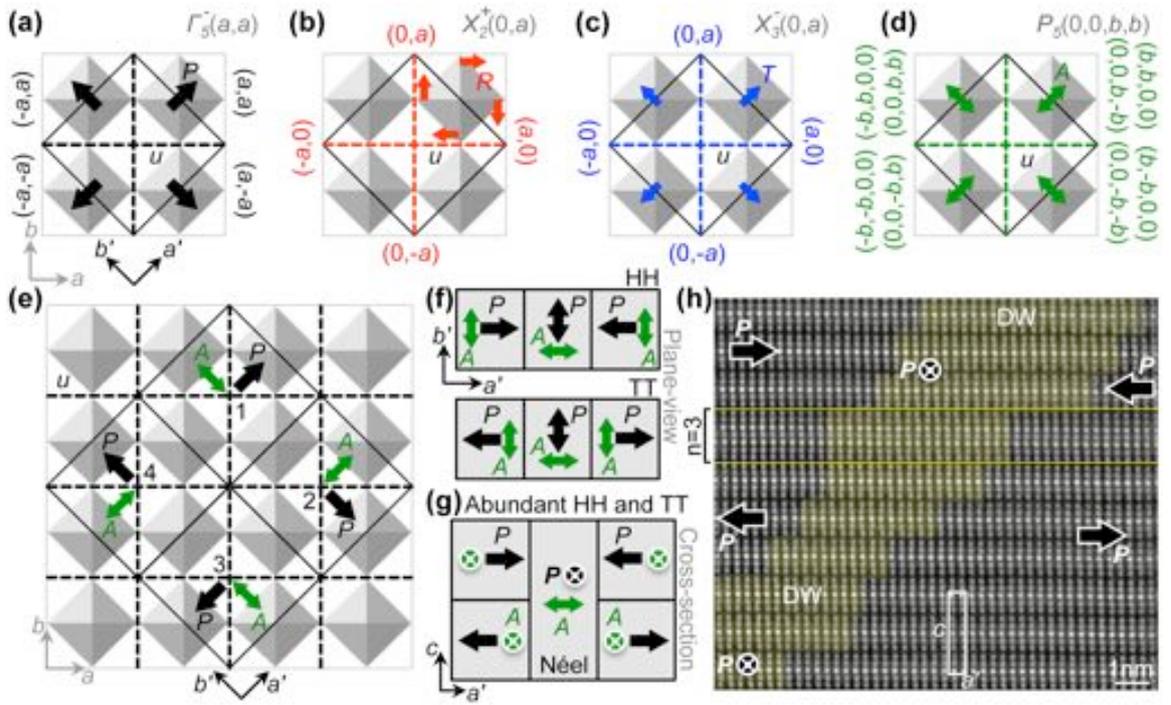

FIG. 4 (M. H. Lee *et al.*)

# Supplemental Material
# Hidden Antipolar Order Parameter and Entangled Néel-Type Charged Domain Walls in Hybrid Improper Ferroelectrics

M. H. Lee, C.-P. Chang, F.-T. Huang, G. Y. Guo, B. Gao, C. H. Chen, S.-W. Cheong, and M.-W. Chu

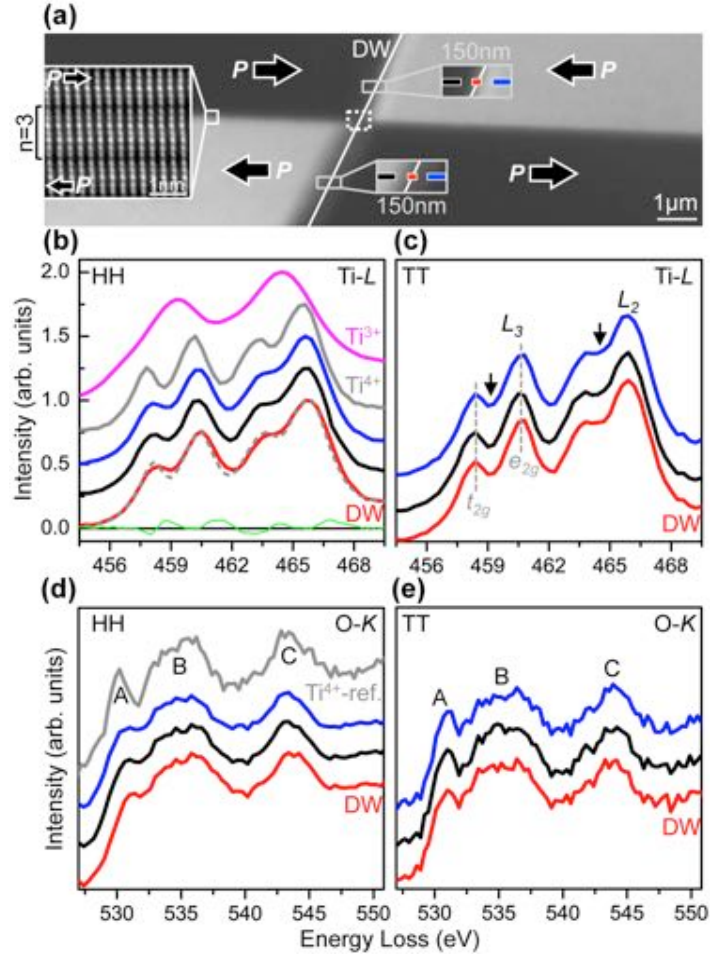

FIG. S1. (a) DF imaging of coexisting HH and TT domains, with the DW regions for STEM-EELS tackling being enlarged for clarity (length, 150 nm). Red (black and blue), the probing across the DWs over 20 nm with 0.5 nm interval per spectrum (of the domains over 40 nm, 1 nm interval). The neighboring HH and TT domains are spaced by intergrowth defects (n = 3, inset). The HAADF imaging corresponding to the dashed rectangle is shown in Fig. 4h (main text). (b) and (c) Ti-$L$ and (d) and (e) O-$K$ edge EELS spectra across the DW (red) and within individual domains (black and blue), with each spectrum being the sum over the 40 ones that show similar characteristics. Gray (magenta) in (b), the $Ti^{4+}$ ($Ti^{3+}$) reference of an as-grown CSTO ($Ti_2O_3$ powders) and summed over 5 ones of comparable quality. Gray in (d), the corresponding O-$K$ edge to the $Ti^{4+}$ reference. Spectra are vertically shifted for clarity. Green in (b), the residual between the HH-DW spectrum (red) and the least-square fit (dashed gray) comprising ~90% $Ti^{4+}$ and ~10% $Ti^{3+}$. In (c), the dips (arrows) of the Ti-$L$ edge spectra are characteristic of $Ti^{4+}$ (gray, (b) panel).

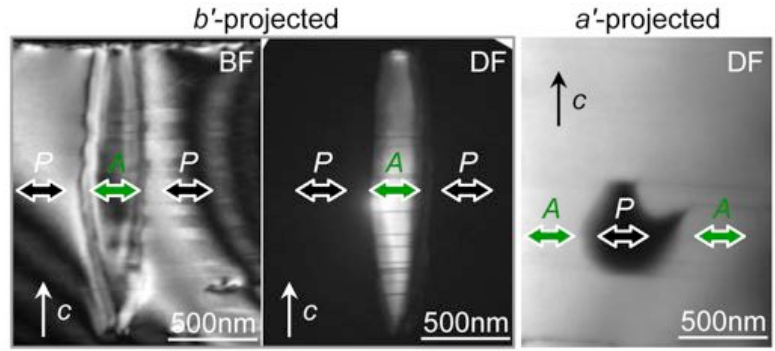

FIG. S2. Irregularly-shaped and large antipolar *A* (polar *P*) domains embedded in *P* (*A*) domains constantly observed in CSTO, affirming the *P-A* intertwined domain topology in Figs. 4e-h (main text). The polarization directions are indistinguishable in these images, thus denoted as doubled-headed *P*.

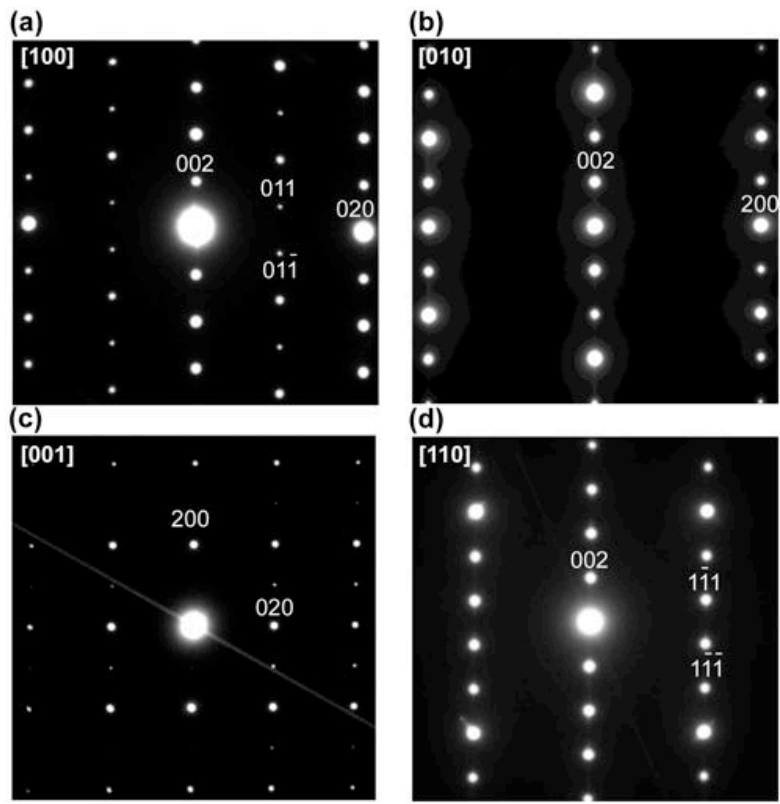

FIG. S3. Selected-area electron diffraction (SAED) patterns of CSTO along (a) 100, (b) 010, (c) 001, and (d) 110 projections in $A2_1am$.

# 1. STEM-EELS probing across HH and TT DWs

The CSTO single crystals were grown by optical floating-zone method and subject to $N_2$ oxygen-deficiency annealing (i.e., electron doping) for enhancing the overall conductivity [18]. Otherwise, the as-grown CSTO crystals are subject to notable charging upon electron-beam illumination, rendering our TEM, STEM, and STEM-EELS tackling unstable. The previous TEM and scanning probe microscopy studies of the CSTO established that both the HH-and-TT coexistence and ground-state structure (orthorhombic $A2_1am$) are robust to the $N_2$ annealing [18-20].

Due to the characteristically small ferroelastic strain of ~0.2% in CSTO, $(a'-b'/a')$ [32], the strain accommodation across the HH and TT DWs can be flexible and, therefore, the DWs would display various angles in addition to 45° (red and blue, Fig. 2d), which is the typical angle for most effectively mitigating noticeable strains [23,32]. With such a small ferroelastic strain, the associated elastic-energy cost of the DWs would also be small.

In Fig. S1, we show the STEM-EELS tackling of charge distributions across the HH and TT DWs. The investigated regions were enlarged for clarity and 40 spectra have been acquired in each individual domain (across DWs) as denoted by the black and blue (red) lines in Fig. S1a. The sums over the 40 spectra, each similar to the others, are exhibited in Figs. S1b and S1c for the Ti-$L$ edge and Figs. S1d and S1e for the O-$K$ edge.

The Ti $L$-edge spectra across the HH (Fig. S1b) and TT DWs (Fig. S1c) are different, showing the dips characteristic of $Ti^{4+}$ basically in the TT spectra (arrows, Fig. S1c). In the Ti $L$-edge spectra of the HH counterparts (Fig. S1b), the noticeably smeared valleys suggest the presence of $Ti^{3+}$ [13] due to the oxygen-vacancy electron doping [18]. Using the reference $Ti^{3+}$ and $Ti^{4+}$ spectra in Fig. S1b (magenta and gray, respectively), we performed a linear least-square fit on the HH-DW's (red) for estimating the $Ti^{3+}$ fraction [13] and obtained ~10% $Ti^{3+}$ (dashed gray, fitted spectrum) with satisfactorily minimal residual (green). The linear fitting on the HH-domain spectra (black and blue, Fig. S1b) gave similar results of ~10% $Ti^{3+}$, corresponding to the oxygen deficiency of ~0.1 and the electron doping level of ~$3.44 \times 10^{20}$ cm$^{-3}$ ≈ ~$4.9 \times 10^{13}$ cm$^{-2}$ [13]. An electrostatic screening of the HH polarizations primitively requires the electron density of $2P/e$ ($e$, elementary electron charge) [13,23-25], leading to ~$3.7 \times 10^{13}$ cm$^{-2}$ using $P$ of ~2.97 μC/cm$^2$ in the single crystals [18]. These comparable electron densities suggest that the $N_2$-annealing electron doping intricately meets the primitive electrostatic boundary condition of HH due perhaps to thermodynamics. Moreover, the doped electrons spread over ~150 nm across the HH DW (Figs. S1a and S1b). However, an electron accumulation within the few-nm wide DW (Figs. 2a, 2e, and 4h; main text) is expected in a classical electrostatic screening context [13,23-25,33]. The characteristic electrostatic screening length at the HH DWs in CSTO was then scrutinized using the corresponding Thomas-Fermi screening formula appropriate for perovskites [33]. Considering the polarization of ~2.97 μC/cm$^2$ and dielectric constant of ~30 for CSTO [18], we obtained the screening length of ~8.1 Å across the HH DWs. With the few-nm DWs resolved in Figs. 2a and 4h, this estimated screening length does correspond to a carrier confinement at the DWs, which is nonetheless discarded by the observed electron spreading over ~150 nm (Figs. S1a and S1b). The smearing out of the dip between O-*K* peaks A and B (Fig. S1d), characteristic of the presence of oxygen vacancies [13,33], agrees with the oxygen-vacancy doped electrons therein (Fig. S1b), and the TT counterparts (Fig. S1e) marginally harbor oxygen vacancies in consistence with the marginal Ti-valence variation across the TT DW (Fig. S1c; our spectroscopic detection limit, 2~3% [13]). It is noted that TT DWs shall be screened by oxygen vacancies in an electrostatic-screening context [27,34]. The appreciable electron spreading across the HH DW (~150 nm, Fig. S1b) and the absence of oxygen vacancies in the neighborhood of TT DW (Fig. S1e) unambiguously discount any electrostatic-screening essence for the coexisting HH and TT DWs. The electron accommodation in HH domains (Figs. S1b and S1d) could be a result of finite residual depolarization fields thereby that point away from the DW and facilitate an electron reservoir, with the inversed fields in the TT domains readily depleting electrons (Figs. S1c and S1e) [13,24,25,33].

## 2. *P-A* intertwined domains and FE switching

We constantly observed large and arbitrarily-shaped antipolar *A* (polar *P*) domains embedded in *P* (*A*) domains such as the *b'*-projected micrographs in Fig. S2 (the *a'*-projected one), providing an additional evidence for the generic *P-A* domain pairing in Figs. 4e-h (main text). In addition to Néel-type DWs, there exists the other type of FE Bloch walls, featuring an out-of-plane 90° rotation of the dipole, and it is assisted by the coexistence of DWs with various non-orthogonal angles in the characteristic order-parameter space [35,36], at odds with the domain topology in CSTO (Fig. 4, main text). Indeed, FE Bloch walls have not been observed in our work. In addition, Fig. S3 shows the SAED patterns of CSTO along (a) 100, (b) 010, (c) 001, and (d) 110 projections, with all observed Bragg spots consistent with the reflection conditions of the macroscopic $A2_1am$ symmetry.

The FE switching in HIF represents an otherwise subtlety, with the field-driven flipping of *P* (formerly ascribed to the antiparallel Ca1/Sr1 and Ca2/Sr2 displacements; Fig. 1a, main text) to be assisted by a costly pathway of one-step reversal of octahedral $X_2^+$-mode rotation or $X_3^-$- tilting (energy barrier, 60~90 meV) [30]. The two-step alternatives of antipolar and twin-assisted pathways (energy barrier, both 30~40 meV) were then theoretically proposed, with the former for the two perovskite slabs composing reversed *P* (thus *Pnam* ground state) and the flip of either one accomplishing the switching, and the latter for two consecutive 90°-twin flips [30]. The antipolar notion [30] is visibly different from our present exploration and, throughout our characterizations, the *Pnam* ground state and related domain structures were not observed. In Figs. 2a, 2e, and 4h, the sandwiched antipolar Néel-type DWs between the HH and TT domains, however, mimics a frozen form of the twin-switching pathway [30]. The CTO and CSTO do feature facile FE switching [18] and the *P-A* intertwining (Figs. 4e-h) appears to capture the favored twin solution. We nonetheless stress that all $\Gamma_5^-$-induced displacements (Fig. 1c) contribute to the ferroelectricity instead of the antiparallel Ca1/Sr1 and Ca2/Sr2 only, similar to the case of proper-FE, layered-perovskite Aurivillius phase of $SrBi_2Ta_2O_9$ (m = 2; m, perovskite-unit number; $A2_1am$) with a switching barrier of 34~35 meV [37,38]. Indeed, the FE switching in Ruddlesden-Popper HIF remains an ongoing subject and a systematic comparison to FE Aurivillius with an identical number of perovskite units would be intriguing.

We are also aware, in Ref. 30, of the consideration of a phase term in the two-dimensional $X_2^+$ and $X_3^-$ irreps in order for tackling the FE switching by $P$ rotation in $a'b'$-plane. Indeed, a phase consideration on such two-dimensional irreps had been exploited in pioneering Ref. 1, however, for capturing the number of domains of a given order parameter through the number of equivalent solutions in the corresponding Landau formula. This pioneering conception [1] forms the basis of the isotropy-subgroups domain topologies in Fig. 4 in the main text [5,39].

## 3. Methods

The BF and DF TEM imaging and SAED were performed on JEOL 2000FX electron microscopy. The CBED was acquired on field-emission FEI Tecnai G2 at 100 K (no further low-temperature phase transition [20,21]), taking advantage of the high spatial coherence of a field-emission gun and the reduced background at low temperatures both essential for quality CBED characterizations [40]. The HAADF STEM imaging and electronic STEM-EELS scrutiny were conducted on spherical-aberration corrected JEOL 2100FX, with the electron-probe size of 0.9~1 Å, the respective HAADF and STEM-EELS collection angles of 70-190 and 30 mrad, and the specimen thickness of 0.4~0.5 λ (λ, inelastic mean-free path) [13]. The HAADF images were Fourier filtered to enhance the contrasts and the spectral integration window of 0.6 eV was exploited in the STEM-EELS mapping [13]. All electron microscopes were operated at 200 keV and we used conventional mechanical polishing and Ar-ion milling to prepare the CSTO specimens [13].

The basis lattice of parent *I4/mmm* was considered in the isotropy subgroups analysis. In addition, the phonon-dispersion calculation was performed on CTO for reducing the calculation cost within the density functional perturbation theory (DFPT) [41] with the generalized gradient approximation (GGA) [42] in *I4/mmm*. The ultrasoft pseudopotential plane wave method implemented in the Quantum Espresso program [43] was used together with a plane wave cutoff energy of 50 Ry. The ground-state electronic structure was calculated self-consistently using a $k$-point mesh of 10×10×5 with an energy convergence criterion of $10^{-6}$ eV. The phonon frequencies and dynamical matrices were calculated within the DFPT for a 4×4×4 $q$-point grid and then Fourier transformed to generate the interatomic force constants. The phonon dispersion and associated modes at any $q$-point in the Brillouin zone were produced by thus-obtained force constants. The *I4/mmm* structure was optimized on the basis of density functional theory with GGA. The full-potential projector-augmented wave method [44], implemented in the Vienna *ab* *initio* simulation (VASP) package [45], was used under the consideration of a large plane cutoff energy of 450 eV and total 54 *k*-points in the irreducible Brillouin zone wedge. Upon the structural optimization, we set the total-energy convergence criterion of $10^{-6}$ eV and all atoms were allowed to relax until the associated forces being smaller than 0.01 eVÅ$^{-1}$. Indeed, the atomic forces from the Quantum Espresso calculations using the VASP optimized structure, which is more computationally efficient, are also small and similar to that from the VASP calculations.